\documentstyle[editedbook,epsfig,psfig,epsf]{mq}

\def\cm2{cm$^2$ }
\def\se1{s$^{-1}$ }

\newcommand{\gx}{{GX~339$-$4}}
\newcommand{\cyg}{{Cyg~X-1}}
\newcommand{\lmc}{{LMC~X-3}}

\begin{opening}
\title{State Transitions in Black Hole Candidates}
\author{M. A. Nowak$^1$}
\institute{$^1$ MIT-CXC, NE80-6077, 77 Massachusetts Ave., Cambridge, MA ~02139, USA}
\end{opening}

\runningtitle{State Transitions in Black Hole Candidates}
\runningauthor{Nowak}

\begin{document}
\vspace{-0.5cm}
\begin{abstract}
{\small As described in the review by Marek Abramowicz (these
proceedings), accretion theory is at something of a crossroads.  Many
theoretical descriptions of observations have centered on the
``hydrodynamic'' approach of the ``standard'' Shakura \& Sunyaev
$\alpha$-disk \cite{shakura:73a}; however, recent MHD simulations of
(non-radiative) accretion flow have called into question the validity
of this approach \cite{krolik:02a,hawley:01a}.  There has been a great
deal of optimism that these simulations give direct insight into
current observations as similar phenomena appear in both; e.g., jets
\cite{hawley:02b,hawley:01c}, rapid high-amplitude fluctuations
\cite{hawley:02a}, etc.  In comparison to real data from black hole
candidates (BHC), however, these similarities are in many ways only
superficial.  In some aspects, simple theories agree with the
observations quite well.  The observed rapid variability is much more
highly structured than found in simulations, and shows interesting
correlations with spectra that have been interpreted in terms of
simple models.  On the other hand, debates have arisen over even the
most basic phenomenological issues concerning, e.g., the geometry and
dominant radiation mechanisms of the accretion flow.  In this article
I present my views of those observational properties of BHC states
that most urgently need to be addressed, and I briefly discuss some of
the models currently being debated.}
\end{abstract}

\section{States, Transitions, \& Hysteresis}  

\begin{figure}[htb]
\centering \psfig{file=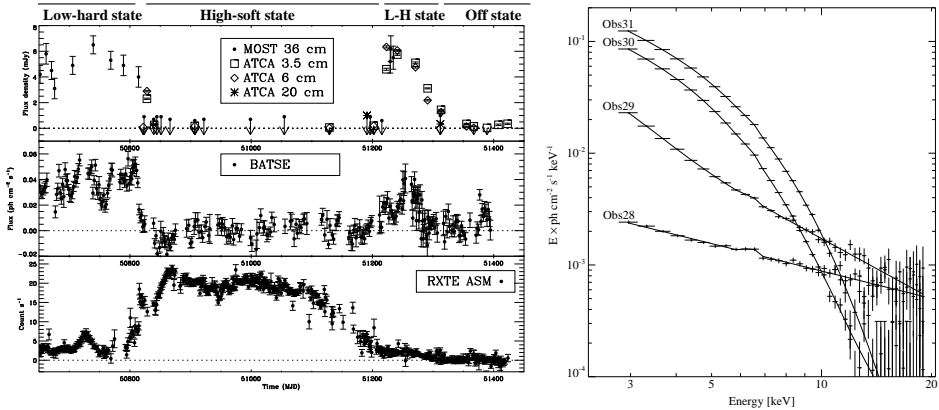,width=13cm}
\caption{{\it Left:} State transitions in the BHC \gx, showing a
simultaneous rise in soft flux with a quenching of the radio
\protect\cite{fender:99b}. RXTE observations of state transitions in
\lmc, showing a clear spectral softening as the flux rises
\protect\cite{wilms:01a}.}
\label{fig:trans}
\end{figure}

BHC states can be broadly classified into two categories: hard with high
variability (typically low luminosity), and soft with weak to moderate
variability (typically high luminosity).  Historically, these have
been classified as ``low'' and ``high'' states, respectively, with
radio observations showing the former to be radio loud, and the latter
to be (mostly) radio quiet (e.g., \cite{fender:99b}). Examples of
transitions between states are shown in Fig.~\ref{fig:trans}.  The
``off'' state appears to be a low luminosity hard state
\cite{kong:00a}, and not necessarily a distinct state.  The ``very
high state'' is a high luminosity soft state wherein the flux of the
hard tail, the X-ray variability, and the radio emission are elevated
from the very low levels observed in moderate luminosity soft states.
Compared to the hard state, the very high state radio emission is more
episodic and its spectrum is steeper, evocative of the ``ejection''
behavior seen in GRS~1915+105 (e.g., \cite{pooley:97a,belloni:97b}).
It is interesting to note that the very high state of \gx, based
solely on the X-ray observations, was suggested to be a jet-ejection
state \cite{miyamoto:91b}.  The ``intermediate state'' is somewhat
ill-defined, and is similar to a very high state (i.e., increased hard
tail and X-ray variability, with variable radio).  This label has been
applied to spectra intermediate between ``high'' and ``very high''
states, as well as to spectra occurring inbetween ``hard/low'' and
``soft/high'' states.  The behavior exhibited by \cyg\ during its
state transitions and ``failed'' transitions \cite{pottschmidt:02a}
may also qualify as an ``intermediate state''.  (The canonical soft
state of \cyg, however, is itself in many ways reminiscent of
definitions of the ``intermediate state''.)  In this review, I will
primarily refer to states as being ``hard'' or ``soft'', rather than
use the labels of off/low/intermediate/high/very high.

X-ray spectra of low variability soft states are in fact remarkably
well-described by simple disk models (with fitted inner disk radii
comparable to the expected marginally stable orbit radii), and do not
show any of the variability seen in MHD simulations.  Nor do they
exhibit the instability behavior normally associated with
$\alpha$-disks \cite{piran:78a}, despite the fact that their
luminosities can be $\approx 5$--$10\%$ of their Eddington luminosity.
Although transitions between extremes of the soft state and extremes
of the hard state appear to track accretion rate, the luminosity at
which these transitions occur is not fixed, and in fact exhibits
evidence of hysteresis \cite{miyamoto:95a}. That is, a state tends to
remain soft to lower luminosity levels, or hard to higher luminosity
levels, if it began in a soft or hard state, respectively.  This is
observed in recent state transitions of \gx\ \cite{nowak:02a}, as well
as for numerous other BHC and neutron star X-ray binaries
(\cite{muno:02a,barret:02a}; Maccarone et al., in prep.).  Very
complex behavior can be seen (e.g., the ``comb-like'' color-intensity
diagrams of XTE J1550$-$564; \cite{homan:01a}), and it is unlikely
that accretion rate, even accounting for hysteresis, is the
sole-determinant of state transition behavior.  As I further discuss
below, even for a given luminosity \emph{and} spectral hardness, it is
not clear that any remaining spectral features (e.g., line strengths
or widths) remain uniform from one instance of a given state to
another.

\section{Theoretical Prejudices}

\begin{figure}[htb]
\centering
\psfig{file=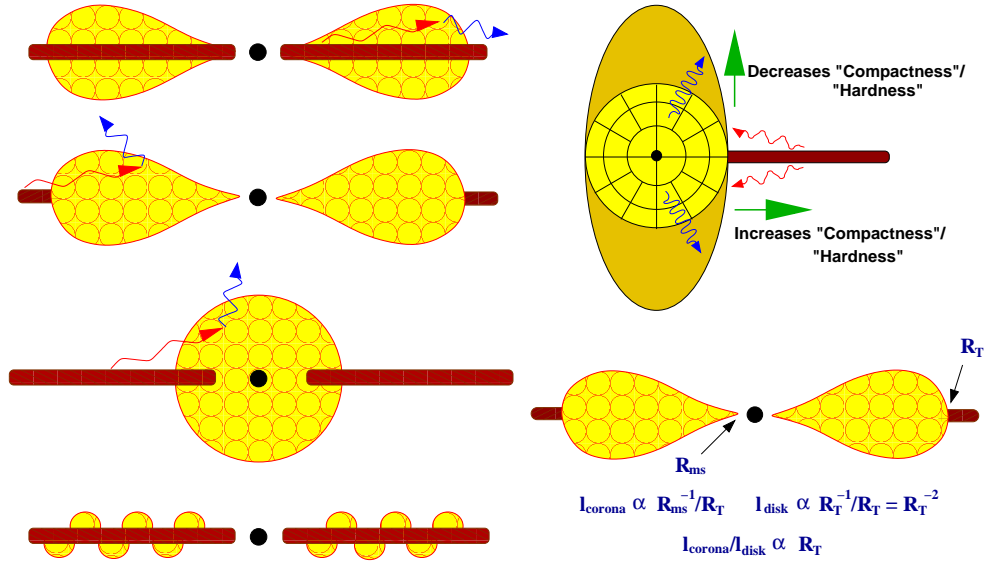,width=13cm}
\caption{{\it Left:} Suggested geometries of the disk and corona for
Comptonization models of the hard/low state. {\it Right:} In lieu of
photon index, $\Gamma$, I will sometimes describe spectral hardness
via a coronal compactness.  I show here how the coronal compactness
changes for variations of a specific geometry.}
\label{fig:combo}
\end{figure}

The observational features that seem to be of the most fundamental
importance in the definitions of BHC states are the presence (or lack
thereof) of hard X-ray and radio emission.  The former has been
appreciated for nearly 30 years; however, the intimate coupling of the
radio and hard X-ray emission is something that has been appreciated
for only the past 6 years or so.  (See the review by Fender in these
proceedings.)  Theoretical models translate the presence/absence of
radio/hard X-ray emission into the presence/absence of either a
Comptonizing corona and/or a synchrotron emitting jet.  The latter
possibility is more thoroughly discussed in the reviews of Falcke and
Markoff (these proceedings; see also \cite{markoff:01a}).  Here I
shall predominantly focus on coronal models, especially as they relate
to models of the hard state.  (It should be noted that a Comptonizing
corona and synchrotron emitting jet are not necessarily mutually
exclusive; however, such ``hybrid'' models are only now beginning to
receive theoretical attention.)

Coronal models are promising in that they can produce naturally both
the power law photon index as well as the sharp cutoff at energies of
$\approx 100$\,keV that are typically observed in hard state BHC.
Debate, however, centers around the geometry of the disk/corona
system.  Several suggested geometries are shown in
Fig.~\ref{fig:combo}.  The ``sandwich geometry'', shown in the upper
left, had been considered by numerous authors \cite{haardt:91a};
however, as 100\% of the downward directed hard photons interact with
the disk, most of which are then thermalized and re-radiated as soft
photons into the corona, it is impossible to create a corona hot
enough to produce sufficiently hard spectra as seen in BHC
\cite{dove:97b}.  This has led researchers to postulate more ``photon
starved'' geometries, such as the bottom three geometries on the
left. In these geometries, a relatively small fraction of the hard
photons intercepted by the disk are re-radiated into the corona as
soft photons.  The degree to which a ``reflection component'' is
generated also varies with these geometries. The bottom left model
(``pill box'' or ``patchy corona'' geometry) yields a large amount of
reflection (all the downward directed photons are intercepted by the
disk), with little soft photon-cooling of the corona \cite{stern:95a}.

The spectral hardness, as represented by the photon index $\Gamma$,
produced within these models is determined by temperature and optical
depth of the corona, as well as by the geometry.  For example, for the
geometry with a spherical corona overlapping an outer disk shown in
Fig.~\ref{fig:combo}, the spectral hardness is partially governed by
the degree of overlap between corona and disk, with greater overlap
yielding softer spectra (e.g.,\cite{zdziarski:99a}).  Instead of
describing spectral hardness via $\Gamma$, one can instead refer to
the coronal compactness, $\ell_c$.  Here I define this as the ratio of
energy generation within the corona divided by its characteristic
radius, to energy generation within the disk divided by its
characteristic radius.  As an example I shall return to below, for the
inner corona/outer disk model shown in Fig.~\ref{fig:combo}, if the
disk \emph{specific energy} goes as the inverse of the disk inner
radius (i.e., gravitational energy), and the coronal specific energy
is a fixed quantity (e.g., gravitational for a fixed inner boundary),
then the compactness scales as the inner disk radius \cite{nowak:02a}.
Higher compactnesses yield harder spectra.  Again, one sees that
variations of the coronal geometry can regulate the spectral hardness.

\section{Correlations}

It is fairly easy for a number of the above coronal models to fit a
single, given observation of a hard state BHC.  More stringent tests
occur when multiple observations of the same, or similar, hard state
BHC are considered.  A number of interesting correlations arise that
must be addressed by any theoretical model of these systems.  Here I
discuss what I consider to be among the most important of these
correlations.

\subsection{Flux/Spectra Correlations}

Over a wide range of hard state flux, higher flux typically means
\emph{softer} spectra, as seen in \gx\ \cite{nowak:02a} and \cyg\
\cite{pottschmidt:02a}.  Such a correlation naturally occurs in
`sphere+disk' coronal models if the inner edge of the disk moves
inward as accretion rate increases.  The corona is more effectively
cooled as the disk moves inwards, which also fits in neatly with
the presumption that the moderate luminosity soft state represents a
``classic'' Shakura-Sunyaev type disk.

As discussed in the reviews of Falcke and Markoff, however, there is a
correlation between the X-ray and radio flux, approximately given by
$F_{X} \propto F_{radio}^{1.4}$, i.e., the radio flux decreases more
slowly than the X-ray flux.  This is not currently accounted for in
coronal models, although it can be accounted for if jet synchrotron
radiation produces both the radio \emph{and} X-ray flux (see the above
cited articles).

\subsection{$\Gamma$:$\Omega/2\pi$ \emph{or} $\Gamma$:$\Delta \Gamma$}

Phenomenologically, a good description of RXTE spectra of hard state
BHC is given by a cutoff broken power law plus Fe line.  The break
energy is at $\approx 10$\,keV, and the power law \emph{hardens} above
the break \cite{wilms:99aa}.  Using more physical models,
these spectra instead can be described by an underlying power law
being reflected from a cold slab, which yields a fluorescent Fe line,
an Fe edge, and a reflected continuum
(e.g.,\cite{matt:93a,magdziarz:95a}).  The degree of reflection is
determined by the solid angle, $\Omega/2\pi$, subtended by the cold
medium as seen by the hard tail.  By considering multiple
observations of multiple hard state objects (including Seyfert
galaxies), it has been claimed that a correlation exists between the
photon index of the underlying power law, $\Gamma$, and the degree of
reflection, $\Omega/2\pi$, with softer spectra showing greater
reflection.  This effect can also be seen with broken power law fits
of multiple observations of a \emph{single} object
\cite{wilms:99aa}. The shape of the spectrum above 10\,keV changes
only slightly while the soft spectral index shows a strong variation,
yielding a positive $\Gamma_{\rm low\,E}$:$\Delta \Gamma$ correlation, where
$\Gamma_{\rm low\,E}$ is the photon index of the low energy power law, and
$\Delta \Gamma = \Gamma_{\rm low\,E}$ -$\Gamma_{\rm high\,E}$
\cite{nowak:02a}.

\begin{figure}[htb]
\centering
\psfig{file=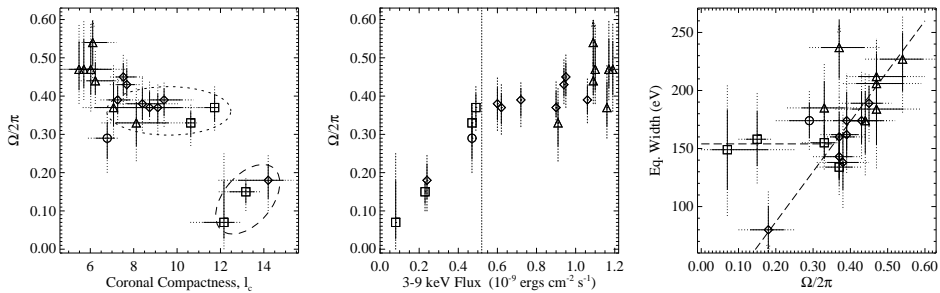,width=13cm}
\vspace{-0.25cm}
\caption{Observations of the hard state of \gx\ \protect\cite{nowak:02a},
showing reflection fraction vs. coronal compactness (left)
and 3--9\,keV flux (middle), and line equivalent width vs. reflection
fraction (right).  Triangles and diamonds are 1997 data, and squares
are 1999 (post-soft state) data.}
\label{fig:reflect}
\end{figure}

The ability of a reflected component to phenomenologically harden a
soft power law has caused some, including myself \cite{wilms:99aa}, to
question whether or not the putative $\Gamma$:$\Omega/2\pi$
correlation is in fact a systematic effect. The correlation, however,
appears robust and does appear in ratios of hard state BHC data to
Crab nebula+pulsar observations (i.e., there is less worry about
systematic uncertainties in the RXTE response matrices, or about the
interdependencies of the model fit parameters; \cite{nowak:02a}).  The
$\Gamma$:$\Omega/2\pi$ correlation is implicitly assigning a
theoretical interpretation, albeit a very plausible one in the context
of some of the coronal models described above.  The purely
phenomenological $\Gamma$:$\Delta \Gamma$ correlation, however, does
need to be addressed by any model of BHC hard states.

Even within the context of reflection, the correlation is not simple.
Fig.~\ref{fig:reflect} shows RXTE observations of \gx\ where the
correlation appears only when considering greatly different
hardnesses.  If one had analyzed only the data with mid-range hardness
circled in Fig.~\ref{fig:reflect}, the correlation would not have been
seen, despite the fact that these observations span a factor of two in
3--9\,keV flux and compactness.  If one instead had analyzed only the
three hardest (and faintest) spectra, \emph{exactly the opposite}
correlation would have been detected.  I note that most of the circled
observations in the mid-range of compactness are hard state spectra in
a high flux regime dominated by ``hysteresis'' \cite{nowak:02a}, and
show a number of other properties that are ``flat'' on various
correlation curves (see Fig.\ref{fig:freqs}).  \gx\ also provides an
example of how spectral properties can apparently change for different
instances of a given state.  Prior to the 1998 soft state, the 1997
hard state showed the Fe line equivalent width to be linearly
correlated with reflection fraction, whereas the 1999 hard state
showed the Fe line equivalent width to be nearly uniform all the way
into quiescence\footnote{It has been recently pointed out that for
RXTE observations, galactic ridge emission can contribute a 6.7\,keV
line with flux of approximately $3\times10^{-5}$ photons/cm$^2$/s
\cite{wardzinski:02a}; however, here the \emph{lowest flux} line has
$2.7\times10^{-4}$ photons/cm$^2$/s, and has a substantial \emph{red
wing} (Fig.~\ref{fig:lines}).  Thus contribution from the galactic
ridge is unlikely to account for the majority of the differences
between the 1997 and 1999 hard states.}  (Fig.~\ref{fig:lines};
\cite{nowak:02a}).

\subsection{$\Gamma$:$f$:$\Delta t$}

\begin{figure}[htb]
\centering
\psfig{file=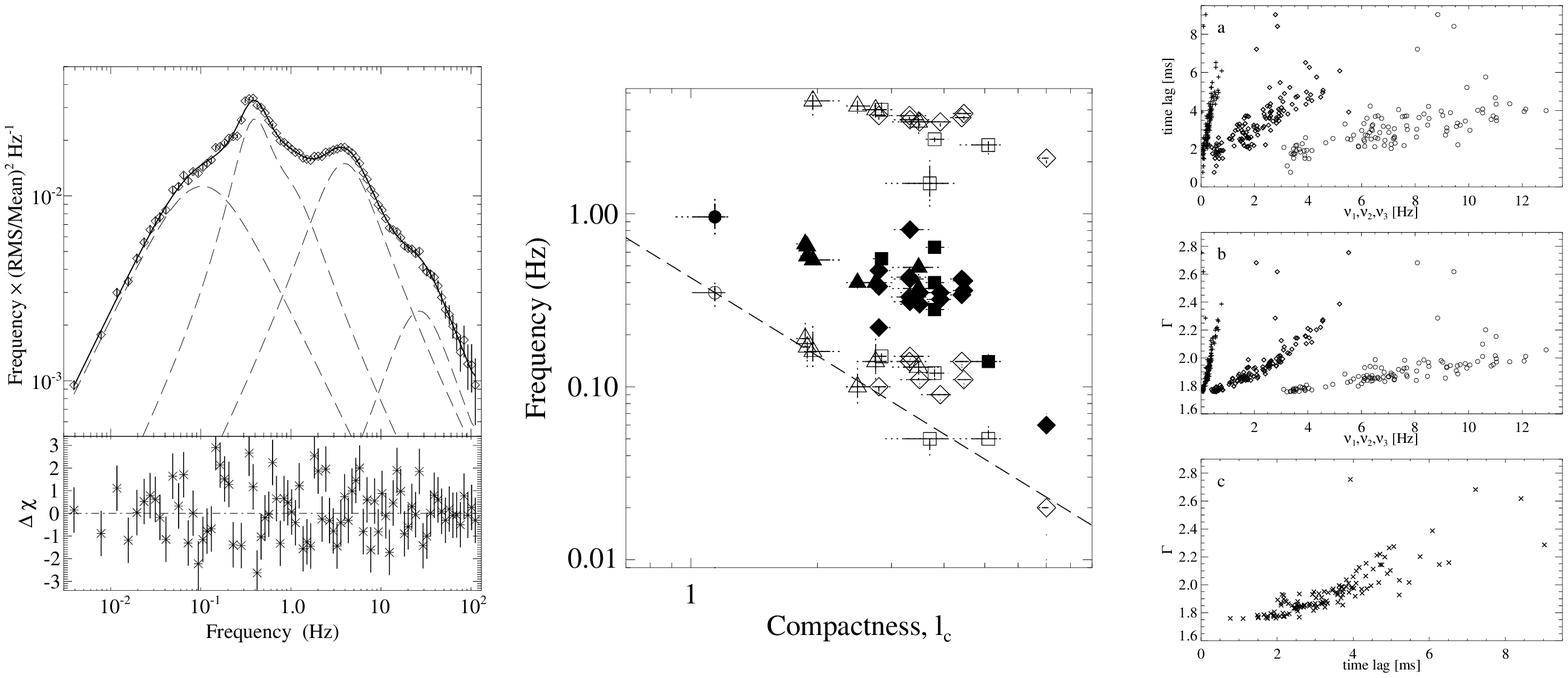,width=12.8cm}
\vspace{-0.3cm}
\caption{{\it Left:} Example of a hard state PSD (\gx,
\protect\cite{nowak:00a}) fit with multiple Lorentzian components.
{\it Middle:} Scaling of these frequency components with coronal
compactness \protect\cite{nowak:02a}.  {\it Right:} Hard
state data of \cyg\ \protect\cite{pottschmidt:02a}, showing photon
index vs. characteristic PSD frequencies (middle), and time lag
between hard and soft X-ray variability (bottom), as well as time lag
vs. frequency (top).}
\label{fig:freqs}
\end{figure}

Correlations also exist between spectra and variability properties,
with harder spectra typically showing \emph{lower frequency} X-ray
variability, both within a given source and across sources
\cite{dimatteo:99a}.  Taking the power spectral density (PSD) of X-ray
variability and multiplying by Fourier frequency, these data often can
be well-described by the sum of broad Lorentzians
(Fig.~\ref{fig:freqs}).  In the hard state, peak frequencies scale as
a function of spectral hardness (Fig.~\ref{fig:freqs};
\cite{dimatteo:99a,nowak:02a,pottschmidt:02a}).  Again, increased
frequency with softer spectra naturally fits in with the concept of
transiting from a radially extended corona to a Shakura-Sunyaev type
disk.  In \gx\, the frequencies approximately scale as $\propto
lc^{-3/2} \propto R_{\rm corona}^{-3/2}$, consistent with naive
expectations of disk time scales being proportional to Keplerian
\cite{nowak:02a}.  Complications with such a simple interpretation
arise, however, when considering time delays between X-ray variability
in soft and hard bands.  Typically, the hard variability lags behind
the soft variability, with this difference \emph{decreasing} for
\emph{lower} Fourier frequency, as shown for RXTE observations of
\cyg\ in Fig.~\ref{fig:freqs} (\cite{pottschmidt:02a}; see also the
paper by Wilms et al., these proceedings).  This is exactly
opposite our naive expectations of time scales increasing with either
the size of our hypothesized corona, or with the characteristic time
scales of the disk/corona transition region.  Any theoretical model
must address the simultaneous correlation of spectral hardness/X-ray
variability frequency/X-ray variability time lags.

\section{The Fender Conjecture}

\begin{figure}[htb]
\centering
\psfig{file=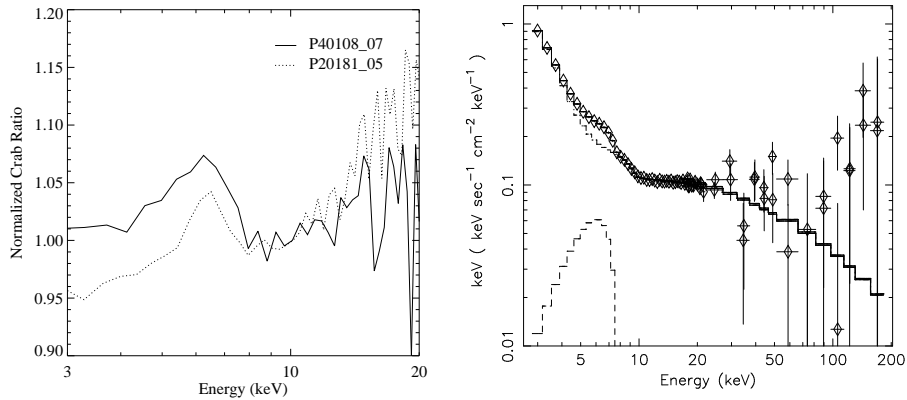,width=12.0cm}
\vspace{-0.3cm}
\caption{Examples of broad lines seen in \gx. {\it Left:} Hard state,
presented as a ratio of RXTE observations of \gx\ divided by RXTE
observations of the Crab nebula and pulsar. {\it Right:} Soft state,
presented as an unfolded RXTE spectrum.}
\label{fig:lines}
\end{figure}

One of the most important questions in black hole research is, what
are the spin parameters of astrophysically observed black holes?
Since the seminal work of Blandford \& Znajek \cite{blandford:77a},
speculation has centered around whether or not rapid (i.e., near
maximal) black hole spin is \emph{required} to produce a jet.  The
seeming ubiquity of radio jets in BHC systems, predominantly
correlated with spectral state (hard spectral states appear to
\emph{always} produce jets), has led to what I shall call the Fender
Conjecture\footnote{The person for whom this conjecture is named in no
way endorses said appellation. I have, however, heard him make this
suggestion in numerous talks; it is an interesting and important
scientific point; and ``The Fender Conjecture'' has at least as much
ring to it as ``The Bourne Identity'' or ``The Corbomite Maneuver''.}:
{\sl If rapid black hole spin is required for producing a jet, then
all black holes are rapidly spinning}.  Clearly, an independent
measurement of spin is required to verify whether or not jetted BHC do
in fact have rapid spin.  One diagnostic that has been suggested is
the relativistic broadening of fluorescent Fe lines \cite{fabian:89a}.

As shown in Fig.~\ref{fig:lines}, broad lines can be found in BHC hard
states.  If such hard states do indeed represent an extended corona
with a truncated disk, however, they do not give the best indication
of spin via their line diagnostics.  Conversely, if soft state spectra
represent disks extending down to, or even within, their marginally
stable orbit, they may offer better line diagnostics.  This,
unfortunately, is not without its own problems.  Specifically, if one
models soft state spectra with a combination of a multi-temperature
blackbody for the soft excess and a power law for any hard tail, these
two fit components tend to cross one another in the Fe line region
(see Fig.~\ref{fig:lines}).  This leads one to worry that any fitted
broad line is in fact a systematic artifact of these fits.  That being
said, a number of soft state BHC spectra do indeed show very strong
and broad lines, so skewed to the red that they are indicative of a
nearly maximally rotating hole.  This can be true even when
Comptonization codes are used to model the hard tail as an upscattered
multi-temperature disk component. (This alleviates some of the
concerns that the broad line is a systematic effect.)
Fig.~\ref{fig:lines} shows such a fit to soft state spectra of \gx\
\cite{nowak:02a}. A clearer example of an extremely broad Fe line,
again consistent with rapid spin, comes from XMM-Newton observations
of the very high state of XTE~J1650$-$500 \cite{miller:02b}. Although
at this point there have been few careful line analyses of soft state
BHC spectra, there is at least a hint of rapid spin in several BHC
systems.

\section{Models and Prospects}

A primary theoretical question is why are there two (broadly defined)
states: soft and hard? As mentioned above, MHD codes show rapid
variability, jets, and magnetic energy dissipation occurring high in
the disk atmosphere \cite{miller:00a}.  Although observations seem to
show more ``regular structure'' (as evidenced by the correlations, the
structured PSD, etc.) than simulations, one still is led to ask why
hard states collapse into structures more quiet, stable, and like a
``standard'' accretion disk as they enter the soft state?  Advection
Dominated Accretion Flow (ADAF) models of hard states
(\cite{esin:97c}, and references therein) only exist below $\approx
10\%$ Eddington luminosity, so they naturally cannot be sustained at
high luminosity.  This, however, does not answer why ADAFs would form
at low luminosity.  Advocates of ADAFs essentially have postulated
what Svensson has dubbed the ``Strong ADAF Principle'': ADAF solutions
are preferred whenever they are physically achievable, and hence
\emph{always} arise at low luminosity.  Taking inspiration instead
from the MHD simulations, Merloni \& Fabian \cite{merloni:02a}
postulate that the rate of magnetic energy deposition into the corona
is strongly dependent upon the \emph{gas pressure}.  Given their
assumptions, they show that the fraction of accretion energy deposited
into the corona scales as $(P_{\rm gas}/P_{\rm total})^{1/4}$. Low
luminosity, gas pressure disks therefore have strong coronae that
collapse as the disk becomes radiation pressure dominated at high
accretion rates.  Two interesting points are raised here: state
transitions might be most intimately tied to whether or not the
accretion flow is radiation pressure dominated; here the
distinction bewtween ``corona'' and ``jet'' is not strong, possibly
leading the way to consideration of hybrid models.

\section*{Acknowledgments}
The author gratefully acknowledges numerous helpful conversations with
O. Blaes, J. Chiang, P. Coppi, S. Corbel, R. Fender, T. Maccarone,
S. Markoff, J. Miller, K. Pottschmidt, C. Reynolds, and J. Wilms, as
well as support from the MIT-Chandra X-ray Science Center via NASA
Grant SV1-61010.


\end{document}